\pdfoutput=1
\documentclass[
 amsmath,amssymb,
 nofootinbib,
 aps,
 prd,
 superscriptaddress,
 floatfix,
]{revtex4}

\usepackage{graphicx}% Include figure files
\usepackage{dcolumn}% Align table columns on decimal point
\usepackage{bm}% bold math
\usepackage{soul} %strike out the text
\usepackage{verbatim}
\usepackage{xcolor,afterpage}
\usepackage{caption}
\begin{document}
\title{The detection of the imprint of filaments on Cosmic Microwave Background (CMB) lensing}

\author{Siyu He}
\affiliation{Carnegie Mellon University, 5000 Forbes Avenue, Pittsburgh PA 15213, USA}
\affiliation{McWilliams Center for Cosmology, Carnegie Mellon University, Pittsburgh, PA 15213, USA}
\affiliation{Lawrence Berkeley National Lab, Berkeley, CA 94720, USA} 

\author{Shadab Alam}
\affiliation{Institute for Astronomy, University of Edinburgh}
\affiliation{Royal Observatory, Blackford Hill, Edinburgh, EH9 3HJ , UK}

\author{Simone Ferraro}
\affiliation{Berkeley Center for Cosmological Physics, University of California, Berkeley CA 94720, USA}
\affiliation{Miller Institute for Basic Research in Science, University of California, Berkeley CA 94720, USA}

\author{Yen-Chi Chen}
\affiliation{Department of Statistics, University of Washington, Seattle, WA 98195, USA}

\author{Shirley Ho}
\affiliation{Carnegie Mellon University, 5000 Forbes Avenue, Pittsburgh PA 15213, USA}
\affiliation{McWilliams Center for Cosmology, Carnegie Mellon University, Pittsburgh, PA 15213, USA}
\affiliation{Lawrence Berkeley National Lab, Berkeley, CA 94720, USA}
\affiliation{Berkeley Center for Cosmological Physics, University of California, Berkeley CA 94720, USA}

\date{\today}

\maketitle
%%%%%%%%%%%%%%%%%%%%%%%%%%%%%%%%%%%%%%%%%%%%%%%%%%%%%
%Intro
\textbf{Galaxy redshift surveys, such as 2dF~\cite{2dF}, SDSS~\cite{SDSS}, 6df~\cite{6df}, GAMA~\cite{GAMA} and VIPERS~\cite{VIPERS}, have shown that the spatial distribution of matter forms a rich web, known as the cosmic web~\cite{bond_theo}. The majority of galaxy survey analyses measure the amplitude of galaxy clustering as a function of scale, ignoring information beyond a small number of summary statistics. Since the matter density field becomes highly non-Gaussian as structure evolves under gravity, we expect other statistical descriptions of the field to provide us with additional information. One way to study the non-Gaussianity is to study filaments, which evolve non-linearly from the initial density fluctuations produced in the primordial Universe. In our study, we report the first detection of CMB (Cosmic Microwave Background) lensing by filaments and we apply a null test to confirm our detection. Furthermore, we propose a phenomenological model to interpret the detected signal and we measure how filaments trace the matter distribution on large scales through filament bias, which we measure to be around 1.5. Our study provides a new scope to understand the environmental dependence of galaxy formation. In the future, the joint analysis of lensing and Sunyaev-Zel'dovich observations might reveal the properties of ``missing baryons'', the vast majority of the gas which resides in the intergalactic medium and has so far evaded most observations.}
\par The cross-correlations of CMB lensing with tracers of large-scale structure have been widely studied\cite{Smith:2007rg,Hirata:2008cb,Bleem:2012gm,Sherwin:2012mr,Ferraro:2014msa,Allison:2015fac,Giannantonio:2015ahz,Eg,Doux:2016xhg,Singh:2016xey,NatAst:1-795}. In our study, we detect the imprint of filaments on CMB lensing by cross-correlating filaments with the CMB lensing convergence ($\kappa$) map. We use the filament intensity map, which is derived from the Cosmic Web Reconstruction filament catalogue~\cite{fil_cat} (Public in https://sites.google.com/site/yenchicr/catalogue) from the Sloan Digital Sky Survey (SDSS)~\cite{SDSS} Baryon Oscillations Spectroscopic Survey (BOSS)~\cite{BOSS} Data Release 12 (DR 12)~\cite{DR}. The filament finder (See Filament Finder in the Method section) partitions the universe from $z =$ 0.005 to $z=$ 0.700 into slices with $\Delta z =$ 0.005. In our study, we use the filaments from $z=$ 0.450 to $z=$0.700, which are detected from CMASS galaxy survey (a galaxy sample from SDSS which targets high redshift). Filaments are found in each redshift bin as the density ridge of the smoothed galaxy density field~\cite{yenchiSCMS} and the filament uncertainty, which describes the uncertainty of filament position, is also calculated (see Uncertainty of Filaments in Method section). 
The filament intensity, illustrated in Supplementary Figure 1, is defined as
\begin{equation}\label{eq:fil_intensity}
I(\hat{n},z) = \frac{1}{\sqrt{2\pi \rho_f(\hat{n},z)^2}}\exp\left(-\frac{\|\hat{n}-\hat{\Pi}_f(\hat{n},z)\|^2}{2\rho_f(\hat{n},z)^2}\right)
\end{equation}
where $\hat{n}$ is the angular position, $\hat{\Pi}_f(\hat{n},z)$ is the angular position of the closest point to $\hat{n}$ on the nearest filament and $\rho_f(\hat{n},z)$ is the uncertainty of the filament at the projected position $\hat{\Pi}_f(\hat{n},z)$.
Using the intensity map at each redshift bin, 
we construct the filament intensity overdensity map via
\begin{equation}
\delta_f (\hat{n}) = \frac{\int I(\hat{n},z)dz - \bar{I}}{\bar{I}},\quad 
\bar{I} = \frac{\int I(\hat{n}, z) \ d \Omega_{\hat{n}} dz}{\int d \Omega_{\hat{n}}}
\end{equation}\\
In this work we use the CMB lensing convergence map (Public in http://pla.esac.esa.int/pla/\#cosmology) from the Planck~\cite{Planck_overview} satellite experiment.  The Planck mission has reconstructed the lensing potential of the CMB from a foreground-cleaned map synthesized from the Planck 2015 full-mission frequency maps using the SMICA code~\cite{Planck}. The lensing convergence $\kappa$ is defined in terms of the lensing potential $\phi$ as 
\begin{equation}
    \kappa(\hat{n}) = \frac{1}{2}\nabla_{\hat{n}}^2\phi(\hat{n}) 
\end{equation}
We measure the cross angular power spectrum of CMB lensing convergence and filaments $C_l^{\kappa f} $ using standard techniques (See Estimator in Method section). We compute the error for each power spectrum by jackknife resampling the observed area into 77 equally weighted regions (see Supplementary Section 1 and Supplementary Figure 2) that comprise the CMASS galaxy survey from where the filaments are detected.  
\par We construct a phenomenological model to describe the cross-correlation of filaments and the CMB lensing convergence field. Instead of modeling the filament profile on small scales\cite{fil_lensing_clampitt,2017MNRAS.468.2605E,Higuchi:2014eia}, our model studies how filaments trace matter distribution on large scales through the use of the filament bias. We assume a $\Lambda$CDM cosmology with Planck parameters from the 2013 release~\cite{Planck_parameter}, where $\Omega_m = 0.315, h = 0.673, \sigma_8 = 0.829, n_s = 0.9603$. In a spatially flat Friedmann-Robertson-Walker universe described by general relativity, the convergence field is
\begin{equation}\label{eq:kappa}
\kappa (\hat{n}) = 4\pi G_N \bar{\rho}_0 \int_{0}^{\chi_{CMB}} \frac{\chi (\chi_{CMB}-\chi)}{\chi_{CMB}}(1+z)\delta_m(\chi,\hat{n})d \chi 
\end{equation}
where $\chi$ is the comoving radial distance, $z$ is the redshift observed at radial distance, $G_N$ is Newton's gravitational constant, $\bar{\rho}_0$ is the present-day mean density of the universe, and $\chi_{CMB}$ is the comoving distance to the CMB. On linear scale, we assume that filaments trace the matter as $\delta_f = b_f \delta_m$, where $b_f$ is defined as the large-scale filament bias. 

 On large scale, we expect the filament overdensity $\delta_f$ to be related to the matter fluctuations through a linear filament bias $b_f$:
\begin{equation}\label{eq:delta_f}
\delta_{f}(\hat{n}) = \int b_f f(z) \delta_m(\hat{n},z)d z
\end{equation}
where $f(z)$ is the mean filament intensity redshift distribution defined as
\begin{equation}
f(z) = \frac{F(z)}{\int F(z)d z},\quad
F(z)=\frac{1}{\Delta z} \int I(\hat{n},z) d\Omega_{\hat{n}} 
\end{equation}
where $I(\hat{n},z)$ is the filament intensity defined in eq.~(\ref{eq:fil_intensity}) and $\Delta z$ is the width of redshift slice. In cross-correlation, on scales smaller than the typical filament length, using filaments introduces additional smoothing compared to the true matter density. We model the smoothing as follows: the filaments have typical length and we lose all small-scale information about fluctuations along the filament; therefore, we take the corresponding filament power spectrum to be exponentially suppressed below the filament scale $k_{\parallel}$ $\sim$ 1/(filament length) in Fourier space. Similarly, any matter in between filaments is either assigned to a filament or eliminated from the catalog (in underdense regions) .  For this reason we also introduce a suppression in power in the direction perpendicular to the filaments, with suppression scale $k_{\perp}$. We use two ways to model $k_{\perp}$. The detailed models are shown later in the paper. Using the Limber approximation~\cite{1953ApJ...117..134L} and the smoothing scale for small scales, the filament-convergence cross-correlation can be written as
\begin{equation} \label{eq:Cl_theo}
C_l^{\kappa f} = \frac{3H_0^2\Omega_{m,0}}{2c^2}\int_{z_1}^{z_2}dzW(z)f(z) \chi^{-2}(z) (1+z) P_{mf} \left(\frac{l}{\chi(z)},z \right)
\end{equation}
where $W(z)=\chi(z)(1-\frac{\chi(z)}{\chi_{CMB}})$ is the CMB lensing kernel,  and $P_{mf}$ is modeled as
\begin{equation}\label{eq:pk_eff}
P_{mf}(k,z) = \frac{1}{2\pi} \int d\phi \ b_f P_{mm}(k,z) e^{-(k \cos(\phi)/k_{\perp}(z))^2-(k \sin(\phi)/k_{\parallel}(z))^2}
\end{equation}
where $P_{mm}$ is the matter power spectrum. We use CAMB (Code for Anisotropies in the Microwave Background) ({\tt http://www.camb.info/}) to evaluate the theoretical matter power spectrum $P_{mm}$. The measurement of filament length is shown in Fig.~1. The mean and median length of the filaments increases as a function of redshift due to the combination of two factors. Firstly, the length of filaments, acting as the mass bridges between galaxy clusters, will decrease. Secondly, the number of filaments detected also depends on the number density of galaxies, which, in the CMASS sample, is low and decreases as a function of redshift (See Supplementary Figure 3). The large difference in the mean and median values of filament length indicates that the distribution of the filament length in each redshift bin is not Gaussian. We plot in the background the 2D histogram of filament length distribution as a function of redshift and filament length. 

\par To check the validity of our model, we also compare the results to simulations. The excellent agreement that we find in simulations provides an important consistency check. The theoretical prediction for $C_l^{\kappa f}$ is shown in eq.~(\ref{eq:Cl_theo}). The matter-filament correlation $C_l^{mf}$ is defined as 
\begin{equation}
C_l^{mf} = \int_{z_1}^{z_2} dz \frac{H(z)}{c} f(z) \chi^{-2}(z) P_{mf} \left( \frac{l}{\chi(z)},z \right)
\end{equation}
By taking the parameters that are slowly varying compared to $f^2(z)$, we get
\begin{equation}{\label{eq:Cl_simu}}
C_l^{\kappa f} = \frac{3H_0^2 \Omega_{m,0}W(z)(1+z)}{2cH(z)} C_l^{mf}
\end{equation}
For the filament catalogue, the effective redshift, defined as the weighted mean redshift of filament intensity, is 0.56.  This approximation is not perfect, leading to a systematic bias in the prediction for $C_l^{\kappa f}$. We propose an estimator for this systematic bias in Supplementary Section 2~\cite{Eg}. As shown in Supplementary Figure 4, the systematic bias is less than 5\%. Thus, the approximation only causes a negligible bias. In our analysis, we measure $C_l^{mf}$ using 10 realizations of sky mocks of dark matter and corresponding filaments (See sky mocks for dark matter and filament in Method section).  
\par Fig.~2a shows the cross angular power spectrum of filaments and the CMB lensing convergence field. We bin our sample into 16 $\ell$ bins. Comparing simulation with data,  we get $\chi^2/d.o.f. = 2.38$ with all 16 data points and $\chi^2/d.o.f. = 1.16$ without the first data point. The deviation of the first data point from the prediction is likely due to cosmic variance given the small sky area ($f_{sky}$ = 0.065) covered by the simulations. 

\par We fit the model of eq.~\ref{eq:Cl_theo} to the data with the filament bias $b_f$ as the fitting parameter. We use two different smoothing methods to find $k_{\perp}$. The first method is to define the perpendicular smoothing scale as the filament spacing, since any scale smaller than the filament spacing is smoothed out. The filament spacing is approximately the filament length. Thus, filament length is the overall smoothing scale for the effective power spectrum in eq.~(\ref{eq:pk_eff}). The result is shown by the red line in Fig.~2. The best $\chi^2$ fit gives $b_f$ = 1.68 $\pm$ 0.334. Since filaments also have width, filament spacing may be an overestimate of the smoothing perpendicular to filaments. In the second model, we also fit for smoothing scale in the perpendicular direction as a free parameter, where we assume $1/k_{\perp}(z) \sim \alpha \times 1/k_{\parallel}(z)$. We get $\alpha$ = 0.65 and $b_f = 1.47 \pm 0.28$. The result is shown as the orange line in Fig.~2.
\par We measure the significance of the cross-correlation detection by measuring the signal-to-noise ratio (SNR). Our SNR is defined as follows 
\begin{equation}
    S/N = \sqrt{\chi^2_{NULL} - \chi^2_{fit}}
\end{equation}
where 
\begin{align}
    &\chi^2_{NULL} = \sum_{ij}d_i^T \left(C_{ij}^{-1} \right) d_j \\
    &\chi^2_{fit} = \sum_{ij}(d_i - t_i)^T \left(C_{ij}^{-1} \right) (d_i - t_i)
\end{align}
where $d_i$ is the cross angular signal in bin $i$, $t_i$ is the best-fit theoretical prediction for the cross signal in bin $i$, and $C$ is the covariance matrix estimated from jackknife resampling. The final result is shown in Table \ref{tab:result}. The SNR values for both models show a significant detection of the cross-correlation. On large scales, we find that the filaments trace the matter with the filament bias around 1.5, which is somewhat smaller than galaxy bias from the same sample.
\par In order to validate the detection of our cross power spectrum, we perform a null test as follows. We rotate the CMB lensing convergence map by $90^{\circ}$, $135^{\circ}$ and $180^{^{\circ}}$, and then we cross correlate these rotated CMB convergence maps with the filament intensity map. Fig.~2b shows that the cross signal with the rotated maps fluctuates around 0. $\chi_{NULL}^2/d.o.f.$ for the three cross angular power spectra is 0.79, 0.75 and 1.04, which means the cross-correlation between rotated CMB maps and the filament intensity map is consistent with 0. In addition, in order to test the impact of lensing generated by the clusters at the intersection of filaments on our signal, we mask out the redMaPPer clusters~\cite{Rykoff:2013ovv} in the CMB lensing map, finding a less than 4\% difference in the cross angular power spectrum. 

\par We define the cross-correlation coefficient between the filament and galaxy maps as $\rho = C_l^{fg}/\sqrt{C_l^{ff}C_l^{gg}}$ , where $C_l^{fg}$ is the cross angular power spectrum of filaments and galaxies, $C_l^{ff}$ and $C_l^{gg}$ are the auto angular power spectrum of filaments and galaxies. The result is shown in the left panel of Fig.~3a. The signal is highly correlated on large scales, since both galaxies and filaments trace the large-scale structure of the matter. However, the correlation decreases on small scales. Fig.~3b shows the cross-correlation of $C_l^{\kappa f}$ and $C_l^{\kappa g}$, where $C_l^{\kappa g}$ is the angular cross power spectrum of the CMB lensing convergence map and the CMASS galaxy catalogue. These two figures show that the maps are not totally correlated with large deviations at small scales. 
Establishing the amount of extra cosmological information present in the filaments field would require a joint analysis with galaxy clustering and lensing measurements; this is left to future work.

\par In our work, we have detected the effect of filaments lensing on the CMB by correlating filaments intensity map with CMB lensing convergence map. We measured filament bias, which is a quantitative description of how filaments trace the underlying matter, to be around 1.5. We perform null tests by rotating the CMB lensing map by more than its correlation length, obtaining results consistent with the null hypothesis. By comparing filaments with galaxies (both at the map and power spectrum level), we show an imperfect correlation, suggesting that there might be additional information in the structure of the cosmic web, of which filaments provide an essential ingredient. In our study, the filament bias measured is significantly different from the mean bias of the CMASS galaxies used to create the filament catalog. This has important consequences for the environmental dependence of galaxy formation and can be key in generating accurate mocks for the next generations of surveys. In addition, the gas in filaments has been recently detected through the thermal Sunyaev-Zel'dovich (tSZ) effect derived from Planck maps~\cite{2017arXiv170910378D} by measuring the gas pressure. A joint analysis of the mass profile and gas pressure can shed light on the majority of the gas in the intergalactic medium that resides outside of halos and hasn't been characterized so far.\\

%%%%%%%%%%%%%%%%%%%%%%%%%%%%%%%%%%%%%%%%%
\noindent \textbf{Contact Information} Correspondence and requests for material should be addressed to S.He.\\
\\
\textbf{Acknowledgements} We thank Anthony Pullen and Elena Giusarma for helpful discussion, Martin White for providing us the N-body simulations, Alex Krolewski, Benjamin Horowitz for comments on the draft. S.Ho is supported by NASA and DOE for this work. S.He is supported by NSF-AST1517593 for this work. S.A. is supported by the European Research Council through the COSFORM Research Grant (\#670193). S.F. thanks the Miller Institute for Basic Research in Science at the University of California, Berkeley for support.  Some of the results in this paper have been derived using the HEALPix package  \cite{2005ApJ...622..759G}.\\
\\
\textbf{Author Contributions} S.He led the project and most of the manuscript writing. S.A. provided the sky mocks for galaxies and dark matter particles as well as wrote the text relative to sky mock for Filaments and Dark Matter in the Method section. S.F. helped with the theoretical modeling and the interpretation of the results, as well as writing part of the manuscript. Y.C. provided the filament intensity maps for data and simulations. S.Ho conceived the idea of cross-correlating filaments with CMB lensing. All authors contributed to the interpretation of the data and commented on the manuscript.\\
\\
\textbf{Author Information}\\ McWilliams Center for Cosmology, Carnegie Mellon University, Pittsburgh, PA 15213, USA\\
Siyu He \& Shirley Ho
\\
McWilliams Center for Cosmology, Carnegie Mellon University, Pittsburgh, PA 15213, USA\\
Siyu He \& Shirley Ho\\
Lawrence Berkeley National Lab, Berkeley, CA 94720, USA\\
Siyu He \& Shirley Ho\\
Institute for Astronomy, University of Edinburgh\\
Shadab Alam\\
Royal Observatory, Blackford Hill, Edinburgh, EH9 3HJ , UK\\
Shadab Alam\\
Berkeley Center for Cosmological Physics, University of California, Berkeley CA 94720, USA\\
Simone Ferraro \& Shirley Ho\\
Berkeley Center for Cosmological Physics, University of California, Berkeley CA 94720, USA\\
Simone Ferraro\\
Department of Statistics, University of Washington, Seattle, WA 98195, USA\\
Yen-Chi Chen\\
%%%%%%%Fig%%%%%%%%%%%%%%%%%%%%%%%%%%%%%%%%%%%%%%%%%
\begin{figure*}
\begin{tabular}{c}
\includegraphics[scale=0.6]{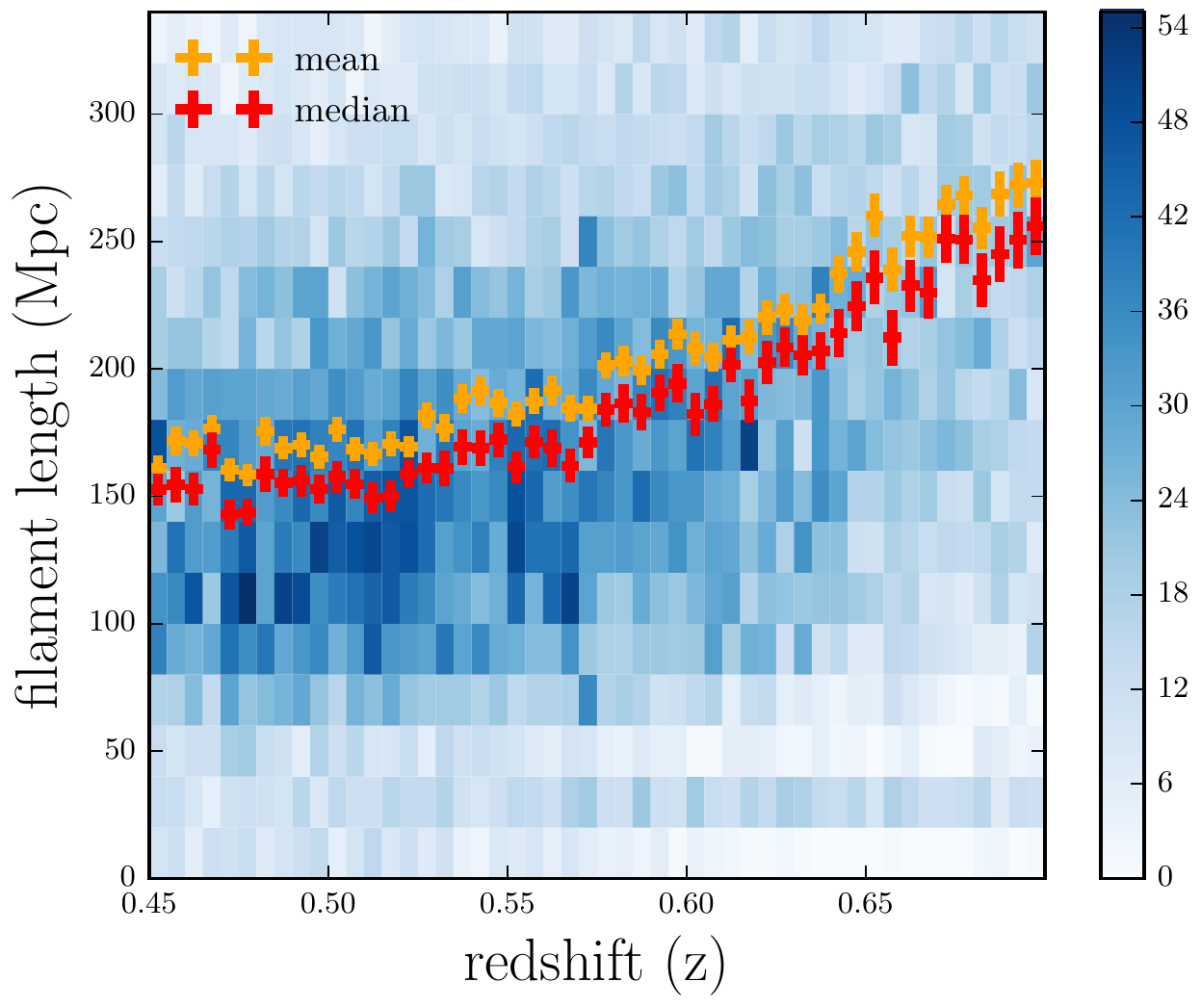}\\
\end{tabular}
\caption{The filament length as a function of redshift. The orange (red) crosses are the mean (median) of the filament length in each redshift bin, where the error bars come from the standard error of the mean (median). The large difference in the mean and the median values implies the filament length distribution is not Gaussian. The background mesh plot shows the 2d histogram of the number of filament length as a function of the redshift and the filament length.}
\label{fig:length}
\end{figure*}

\begin{figure*}
\begin{tabular}{c}
\includegraphics[width = 0.9\textwidth]{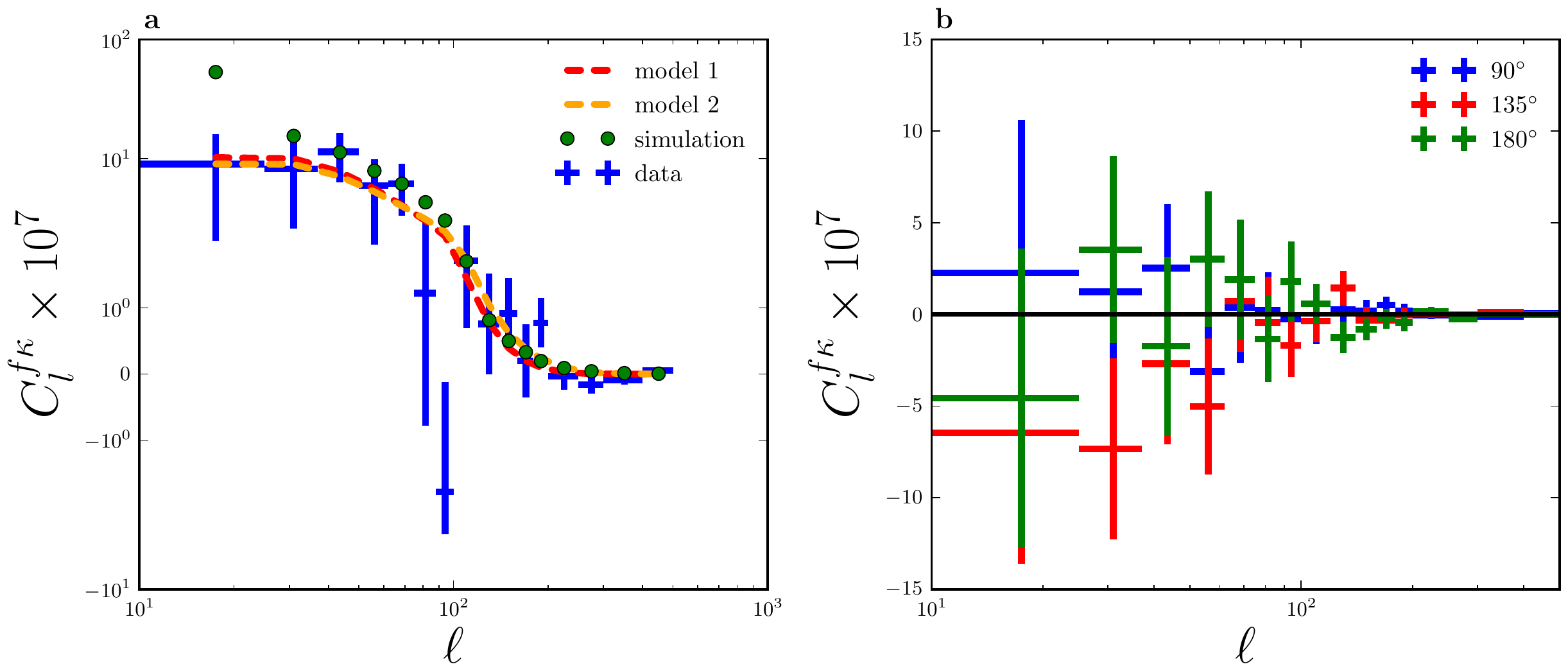}
\end{tabular}
 \caption{Cross angular power spectrum. $(\bm{a})$ shows the cross angular power spectrum of the filaments and the CMB convergence field. The blue crosses are measured with error bars from jackknife resampling of the sky into 77 equally weighted regions. The red and orange dashed lines are theoretical predictions based on different smoothing models (red: filament length and spacing smoothing, orange: filament length and statistical fit for perpendicular smoothing). The corresponding filament bias for the two models are 1.68 and 1.47. The green circles are from simulations. $(\bm{b})$ a null test showing the cross angular power spectrum of the filament catalogue and the rotated CMB lensing convergence map. The cross signals fluctuate around 0. The $\chi^2_{NULL}/d.o.f.$ for the three scenarios are all $\sim$ 1.  }
\label{fig:cl_signal}
\end{figure*}

\newpage

\begin{figure*}[ht]
\begin{tabular}{c}
\includegraphics[width = 0.9\textwidth]{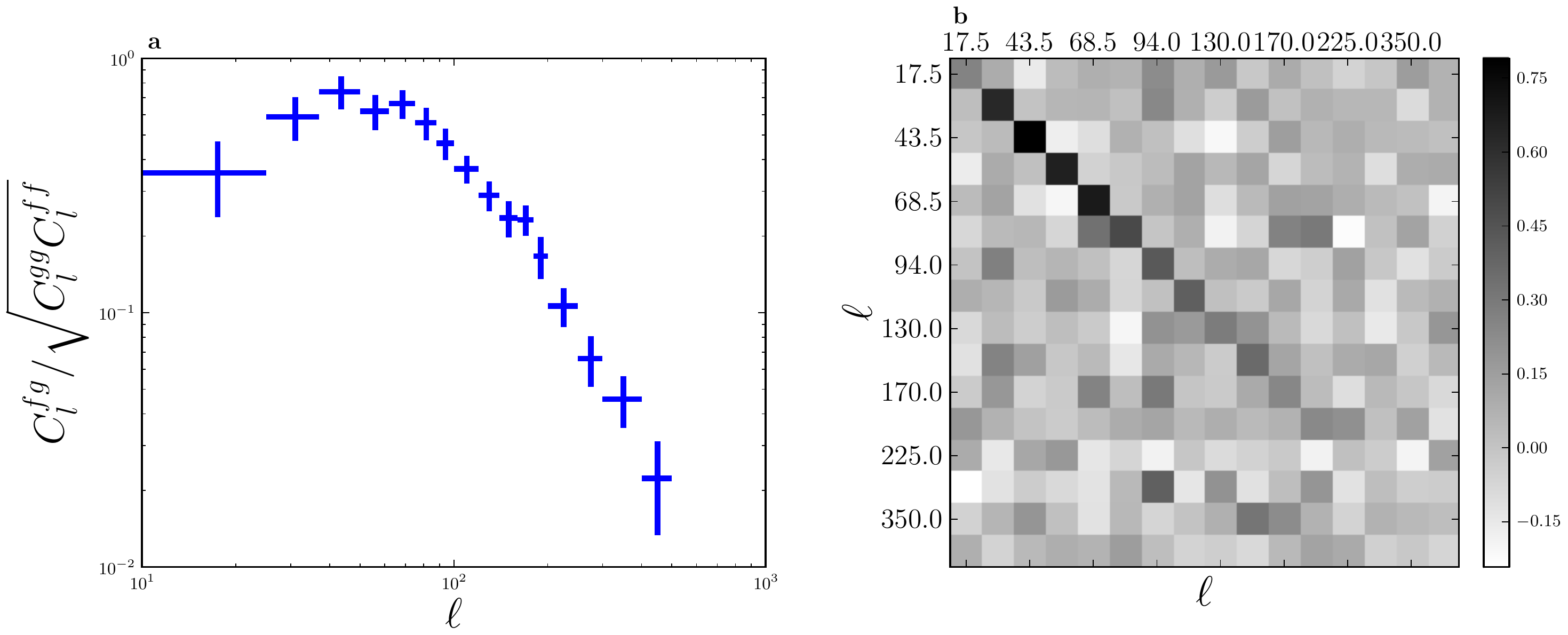}
\end{tabular}
 \caption{ (\textbf{a}) The correlation coefficient of galaxy map and filament map and \textbf{b} the correlation of $C_l^{\kappa f}$ and $C_l^{\kappa g}$. Both plots show the filament and galaxy maps are not totally correlated with large deviations on small scales. }
\label{fig:fil-gal}
\end{figure*}

\begin{table*}[ht]
\scalebox{1.2}{
\begin{tabular}{ |c|c|c|c| }
 \hline
  &model 1  & model 2 \\ 
 \hline \hline
  $b_f$ & 1.68$\pm$0.334 & 1.47$\pm$0.28 \\
 \hline
  $S/N$ & 5.0 & 5.2\\
  \hline
  $\chi^2_{fit}$ & 25.77 & 24.39 \\
  \hline
  $d.o.f.$ & 15 & 14  \\
\hline
\end{tabular}}
\caption{The final result for the bias fitting. Model 1 uses filament length as the overall smoothing scale. In model 2, the filament length is the smoothing along filaments; we fit $\alpha$ for the smoothing in the perpendicular direction, where $1/k_{\perp}(z) \sim \alpha \times 1/k_{\parallel}(z)$. We get $\alpha$ to be 0.65 as the best fit. The bias and the error of bias come from $\chi^2$ fitting of theory model to data.} \label{tab:result}
\end{table*}
%%%%%%%%%%%%%%%%%%%%%%%%%%%%%%%%%%%%%%%%%%%%%%%%%%
\section*{Method}
\textbf{Filament finder.} We obtain filaments from the publicly available Cosmic Web Reconstruction filament catalogue~\cite{fil_cat}. It finds filaments 
by applying the ridges finding algorithm (filament detector) \citep{yenchiSCMS}
to the galaxies in SDSS DR 12, ranging from $z=0.050$ to $z=0.700$. 
The spectroscopic galaxies are used since they give a reliable 
redshift estimate. 
Specifically, the catalogue is constructed using the following steps:
\begin{enumerate}
\item Partition the galaxies in redshift $z=0.050 - 0.700$
into 130 redshift bins such that the bin width is $\Delta z= 0.005$. 
Galaxies within the same bin are projected onto a 2D space. 
\item For each bin:
	\begin{enumerate}
	\item Reconstruct the (2D) galaxy probability density field
	by applying a kernel density estimator (KDE) with smoothing bandwidth chosen by the reference rule in \cite{yenchiSCMS}. 
	\item Compute the root mean square (RMS) of the density field ($\rho_{\sf RMS}$) within the area 150 deg $<$ RA $<$ 200 deg and 5 deg $<$ DEC $<$ 30 deg. 
	\item Remove galaxies in the regions where the probability density field is below a threshold density.
	\item Apply the ridges finding algorithm \citep{yenchiSCMS} to the remaining galaxies.
	\item Apply the galaxy mask to remove filaments outside the region of observations. 
	\end{enumerate}
\end{enumerate}

Here are some remarks on the construction of the catalogue.
\begin{itemize}
\item {\bf The 2D projection.} The universe is sliced
and galaxies are projected onto $2D$ space for several reasons.
First, this enhances the stability of the filament detector. 
Second, this avoids the finger of god effect problem. 
Third, it's easy to compare filaments across different redshifts. 
More detailed discussion can be found in \cite{fil_cat}.
\item {\bf Choice of $\Delta z$.}
The choice of $\Delta z = 0.005$ is to balance the 
estimation bias and the estimation random error.
This is related to the so-called bias-variance tradeoff in statistics \citep{wassermann2006all}. 
If $\Delta z$ is very small, there will be a very limited number of galaxies, so
the filament detector will be unstable. 
On the other hand, if $\Delta z$ is large, 
the bin contains a very wide range of the universe so the filamentary structures may
be washed away when projected onto $2D$ angular space. 
$\Delta z = 0.005$ is an empirical rule we discovered when applying to the SDSS data.

\item {\bf Area selection for calculating $\rho_{\sf RMS}$.}
The specific angular space ($[150,200]\times [5,30] \deg^2$) is selected to compute the RMS of the density field.
The region is chosen because it is a wide region which is almost completely observed in SDSS samples.
The range is large enough so the RMS calculation is stable. 

\item {\bf Thresholding.}
Before applying the ridges finding algorithm, 
galaxies are removed in the low density area. 
This thresholding stabilizes the ridges finding algorithm 
because the algorithm is very sensitive to density fluctuations in low density area. 
\end{itemize}
The filament catalogue is shown to have strong agreement with the redMaPPer Catalogue, since most clusters in the redMaPPer Catalogue lie at the intersection of the filaments in the Cosmic Web Reconstruction filament catalogue, which is predicted by theory. The filament catalogue also has good consistency with the Voronoi model~\cite{yenchiSCMS}.
Furthermore, the effects of filaments on nearby galaxy properties (stellar mass, brightness, age, and orientation) are studied and it shows there is strong correlation of galaxy properties with filament environment, which satisfies theory prediction~\cite{gal_property_fil}. \\
\\
{\bf Uncertainty of filaments.}
The uncertainty of filaments is computed using the bootstrap method \citep{chen2015asymptotic,yenchiSCMS}. 
The filament detector returns a collection of points on filaments, which we call as filament points. For a given redshift bin, 
denote $F_1,\cdots,F_N$ as filament points. The uncertainty of
filament points is computed as follows.
\begin{enumerate}
\item  All galaxies (in one bin) are re-sampled with replacement to generate a new set of galaxies
with the same total number of galaxies. This new set of galaxies is called a bootstrap sample. 
\item Apply the same filament finding algorithm to the galaxies in the bootstrap sample.
This yields a new set of filaments, which are called as the bootstrap filaments.
\item The distance of the filament point to the nearest filament point in the bootstrap filaments is calculated.
Denote $\epsilon_1,\cdots,\epsilon_N$ as the distance for each filament point.
This distance serves as an error measurement for $F_1,\cdots,F_N$.
\item Repeat the above 3 steps $1000$ times ($1000:$ the number of bootstrap replicates). For each filament point, there will be $1000$
error measurements. 
For instance,
the $\ell$-th filament point has $1000$
error values:
$$
\epsilon_\ell^{(1)},\cdots, \epsilon_\ell^{(1000)}.
$$
\item Compute the error (uncertainty) of each filament point by the RMS of the $1000$ error measurements.
Namely, for the $\ell$-th filament point, the error is
$$
\mathcal{E}_\ell = \sqrt{\frac{1}{1000}\sum_{j=1}^{1000} \left(\epsilon_\ell^{(j)}\right)^2}.
$$
\end{enumerate}
The bootstrap procedure measures the uncertainty
due to the randomness of sampling \citep{chen2015asymptotic}.\\
\\
\textbf{Filament length measurement.}
We get the filament intersections from Chen et al (Public in https://sites.google.com/site/yenchicr/catalogue) \cite{fil_cat}. For each redshift bin, we use the hierarchical clustering method~\cite{yenchiSCMS} to determine the number of branches at each intersection. The parameters in the hierarchical clustering are chosen to be the same as \cite{yenchiSCMS}:
\begin{equation}
r_{in} = \frac{2\omega}{3}, r_{out} = 2r_{in}, r_{sep} = (r_{in} + r_{out})/2. 
\end{equation}
where $\omega=\omega(z)$ is the smoothing
bandwidth. At each intersection, we find the nearest point to the intersection point from each branch, and we group the nearest point as the filament point belonging to that filament (See Supplementary Figure 5). Then we keep finding the nearest point to the newly grouped filament to find the next filament point belonging to that branch. 
We stop the loop if the distance between filament points is less than $r_{sep}$ and the distance between a filament point and the other intersection point is larger than  $r_{sep}/2$.\\
\\
\textbf{Estimator.}
We construct the filament map using the HEALPix pixelization with Nside=512. The CMB lensing convergence map is given directly by PLANCK using the  HEALPix pixelization with Nside= 2048. We downgrade the lensing convergence map resolution to Nside=512 to cross-correlate with the filament map. The choice of resolution is consistent with the smoothing applied by the filament finder and is large enough to fully resolve the scales relevant to our cross correlation.

\par We measure the cross angular power spectrum for the filament catalogue and the CMB lensing convergence field using a pseudo-$C_l$ estimator:
\begin{equation}
\hat{C}_l ^{\kappa f}= \frac{1}{(2l+1)f_{sky}^{\kappa f}}\sum_{m=-l}^{l}(\delta_f)_{lm}\kappa _{lm}^*
\end{equation}
where $f_{sky}^{\kappa f}$ is the sky fraction observed by both the filament catalogue and the CMB lensing convergence field, $\kappa_{lm}$ is the spherical harmonic transform of the lensing convergence field, and $(\delta_f)_{lm}$ is the spherical harmonic transform of the filament intensity overdensity. The spherical harmonic transform and $C_l$ are computed using HEALPY. \\
\\
\textbf{Sky mock for filaments and dark matter.}
We use $N$-body simulation runs using the TreePM method \citep{Bagla2002,White2002,Reid14}. We use 10 realizations of this simulation based on the $\Lambda$CDM model with $\Omega_m= 0.292$ and $h=0.69$. Although the parameters of the simulations are slightly different from the Planck cosmological parameters, if we compare the matter power spectrum with the cosmological parameters from the simulations and Planck, the difference is within 2\%. Given the current noise in the data, we believe that this small difference is sub-dominant in our paper. These simulations are in a periodic box of side length 1380$h^{-1}$Mpc and $2048^3$ particles. A friend-of-friend halo catalogue is constructed at an effective redshift of $z=0.55$. This is appropriate for our measurement since the galaxy sample used has effective redshift of 0.57. We use a Halo Occupation Distribution (HOD) \citep{Peacock2000,Seljak2000,Benson2000,White2001,Berlind2002,Cooray2002} to relate the observed clustering of galaxies with halos measured in the $N$-body simulation. We have used the HOD model proposed in \cite{Beutler13} to populate the halo catalogue with galaxies. 
\begin{eqnarray}
\langle N_{\rm cen} \rangle(M) &= \frac{1}{2} \left[ 1+ \mathrm{erf}\left( \frac{\log M- \log M_{\rm min}}{\sigma_{\log M}}\right) \right] \,\nonumber \\
\langle N_{\rm sat} \rangle (M) &= \langle N_{\rm cen} \rangle _M \left( \frac{M}{M_{\rm sat}} \right)^\alpha \exp \left( \frac{-M_{\rm cut}}{M}\right)
\label{eqn:HOD}
\end{eqnarray}
where $\langle N_{\rm cen} \rangle (M)$ is the average number of central galaxies for a given halo mass $M$ and $\langle N_{\rm sat} \rangle (M)$ is the average number of satellite galaxies. We use the HOD parameter set ($M_{\rm min}=9.319 \times 10^{13} M_\odot/h, M_{\rm sat}=6.729 \times 10^{13} M_\odot/h ,\sigma_{\log M}=0.2, \alpha=1.1, M_{\rm cut}=4.749 \times 10^{13} M_\odot/h$) from \cite{Beutler13}. We have populated central galaxies at the center of our halo. The satellite galaxies are populated with radius (distance from central galaxy) distributed out to $r_{200}$ as per the NFW profile; the direction is chosen randomly with a uniform distribution.
\par The sky mocks of dark matter and galaxy are obtained from the simulation box using the method described in \cite{2014MNRAS.437.2594W}. We use publicly available ``MAKE SURVEY" (https://github.com/mockFactory/make\_survey) code to transform a periodic box into the pattern of survey. The first step of this transformation involves a volume remapping of the periodic box to sky coordinates preserving the structure in the simulation. This is achieved by using the publicly available package called ``BoxRemap'' (http://mwhite.berkeley.edu/BoxRemap) \cite{2010ApJS..190..311C}. The BoxRemap defines an efficient volume-preserving, structure-preserving and one-to-one map to transform a periodic cubic box to non-cubical geometry. The non-cubical box is then translated and rotated to cover certain parts of the sky. We then convert the cartesian coordinate to the observed coordinate, which is right ascension, declination and redshift. We down-sample the galaxies with redshift to match the mock redshift with the redshift distribution observed in data. We request the reader refer to \cite{2014MNRAS.437.2594W} for more details.  We then apply the filament detection algorithm to these simulated mocks using the method described in Filament Finder.\\
\\
%\bibliography{ref}{}
%\bibliographystyle{unsrt}
\textbf{Data availability} The data that support the plots within this paper and other
findings of this study are available from the corresponding author upon
reasonable request.\\
\\
\textbf{Competing interests} Authors declare no competing financial interests.

%%%%%%%Fig%%%%%%%%%%%%%%%%%%%%%%%%%%%%%%%%%%%%%%%%%

%\bibliography{ref}{}
%\bibliographystyle{unsrt}

%\end{comment}

\textbf{Acknowledgements} We thank Anthony Pullen and Elena Giusarma for helpful discussion, Martin White for providing us the N-body simulations, Alex Krolewski, Benjamin Horowitz for comments on the draft. S.Ho is supported by NASA and DOE for this work. S.He is supported by NSF-AST1517593 for this work. S.A. is supported by the European Research Council through the COSFORM Research Grant (\#670193). S.F. thanks the Miller Institute for Basic Research in Science at the University of California, Berkeley for support.  Some of the results in this paper have been derived using the HEALPix package  \cite{2005ApJ...622..759G}.

%\section*{Author Contribution}
\textbf{Author Contributions} S.He led the project and most of the manuscript writing. S.A. provided the sky mocks for galaxies and dark matter particles as well as wrote the text relative to sky mock for Filaments and Dark Matter in the Method section. S.F. helped with the theoretical modeling and the interpretation of the results, as well as writing part of the manuscript. Y.C. provided the filament intensity maps for data and simulations. S.Ho conceived the idea of cross-correlating filaments with CMB lensing. All authors contributed to the interpretation of the data and commented on the manuscript.

%\section*{Competing interests}
\textbf{Competing interests} Authors declare no competing financial interests.

\newpage
%\includepdf[pages={1}]{supplement.pdf}
%\newpage
%\includepdf[pages={2}]{supplement.pdf}
%\begin{comment}
\appendix
\section{Supplementary information}
\subsection{Jackknife regions}
To select the jackknife regions, we divide the observed sky in rectangular jackknife regions such that each region has same effective observed area by demanding equal number of randoms. We also tried to make a choice to keep the regions as close to square as possible so we don’t introduce extra scales. We found that 11 $\times$ 7 (RA,DEC) jackknife regions satisfy all the constraints. We show an illustration of jackknife regions in Supplementary Figure 2.

\subsection{Error calculation for $C_l^{\kappa f}$  from simulation}
$C_l^{\kappa f}$ and $C_l^{mf}$ are derived as 
\begin{equation} 
\begin{split}
C_l^{\kappa f} = \frac{3H_0^2\Omega_{m,0}}{2c^2}\int_{z_1}^{z_2}dzW(z)f(z) \chi^{-2}(z) (1+z)\\ 
\times P_{mf} \left(\frac{l}{\chi(z)},z \right)
 \end{split}
\end{equation}
\begin{equation}
C_l^{mf} = \int_{z_1}^{z_2} dz \frac{H(z)}{c} f(z) \chi^{-2}(z) P_{mf} \left( \frac{l}{\chi(z)},z \right)
\end{equation}
By removing the appropriate functions from the integrands, which are slowly varying compared to $f(z)$, the correct expression between $C_l^{\kappa f}$ and $C_l^{mf}$
is
\begin{equation*}
C_l^{\kappa f} = \frac{3H_0^2 \Omega_{m,0}W(z)(1+z)}{2cH(z)} C_l^{mf}
\end{equation*}
However, the approximations required to produce this expression are not perfect, causing the estimation of $C_l^{\kappa f}$ from simulation to slightly deviate from the true value of $C_l^{\kappa f}$. We estimate the deviation $\Gamma$ by relating the theoretical prediction for $C_l^{\kappa f}$ and $C_l^{m f}$ by the following equation
\begin{equation}
\Gamma = \frac{2cH(z)C_l^{\kappa f}}{3H_0^2 \Omega_{m,0}W(z)(1+z) C_l^{m f}}
\end{equation}
The result is shown in Supplementary Figure 1. We see that the $\Gamma$ is less than 5\% from unity, which is much smaller than $\Delta(C_l^{\kappa f})/C_l^{\kappa f}$, where $\Delta(C_l^{\kappa f})$ is the error for $C_l^{\kappa f}$. Thus, the approximation for converting $C_l^{m f}$ to $C_l^{\kappa f}$ only causes a negligible bias. 

\begin{figure}[ht!]
\centering
\begin{tabular}{c}
\includegraphics[scale=0.08]{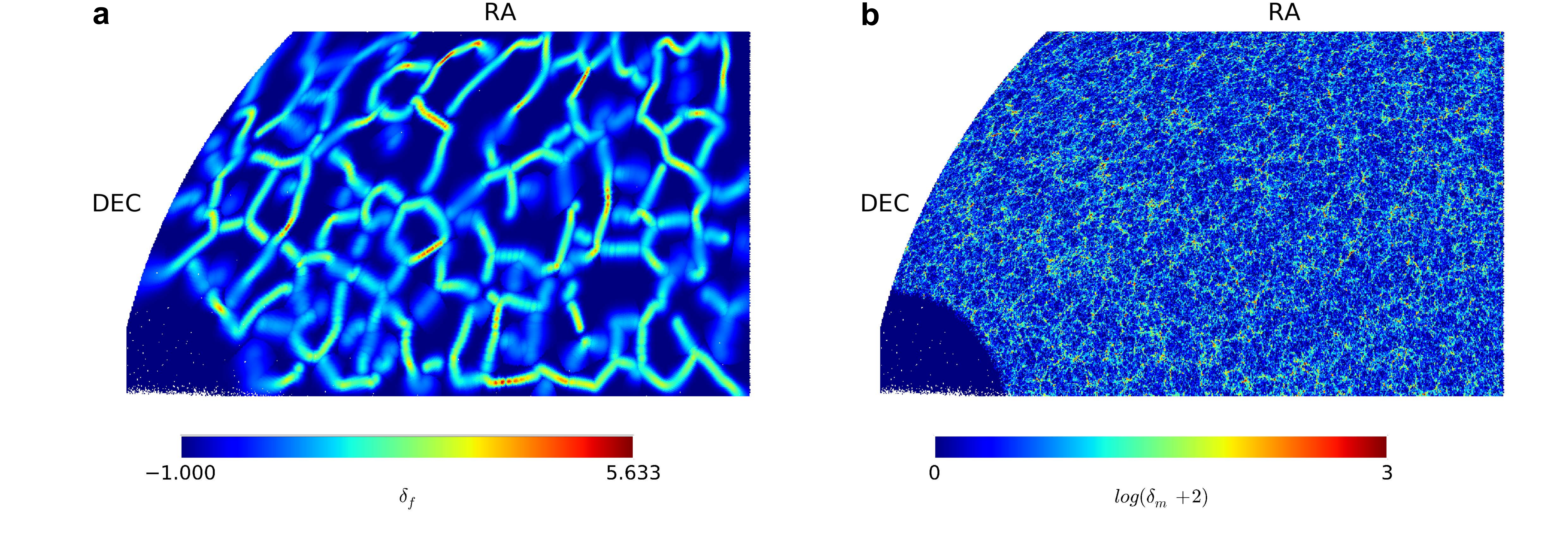}
\end{tabular}
\captionsetup{labelformat=empty}
\caption{\textbf{Supplementary Figure 1:} {\bf A demonstration of filament intensity overdensity and corresponding dark matter particle overdensity in simulation at redshift 0.57.} $(\bm{a})$ filament intensity overdensity at redshift 0.57. $(\bm{b})$ dark matter overdensity at redshift 0.57. The color bars show the amplitude of the overdensity field in linear scale.} 
\end{figure}

\begin{figure}
\centering
\begin{tabular}{c}
\includegraphics[scale=0.7]{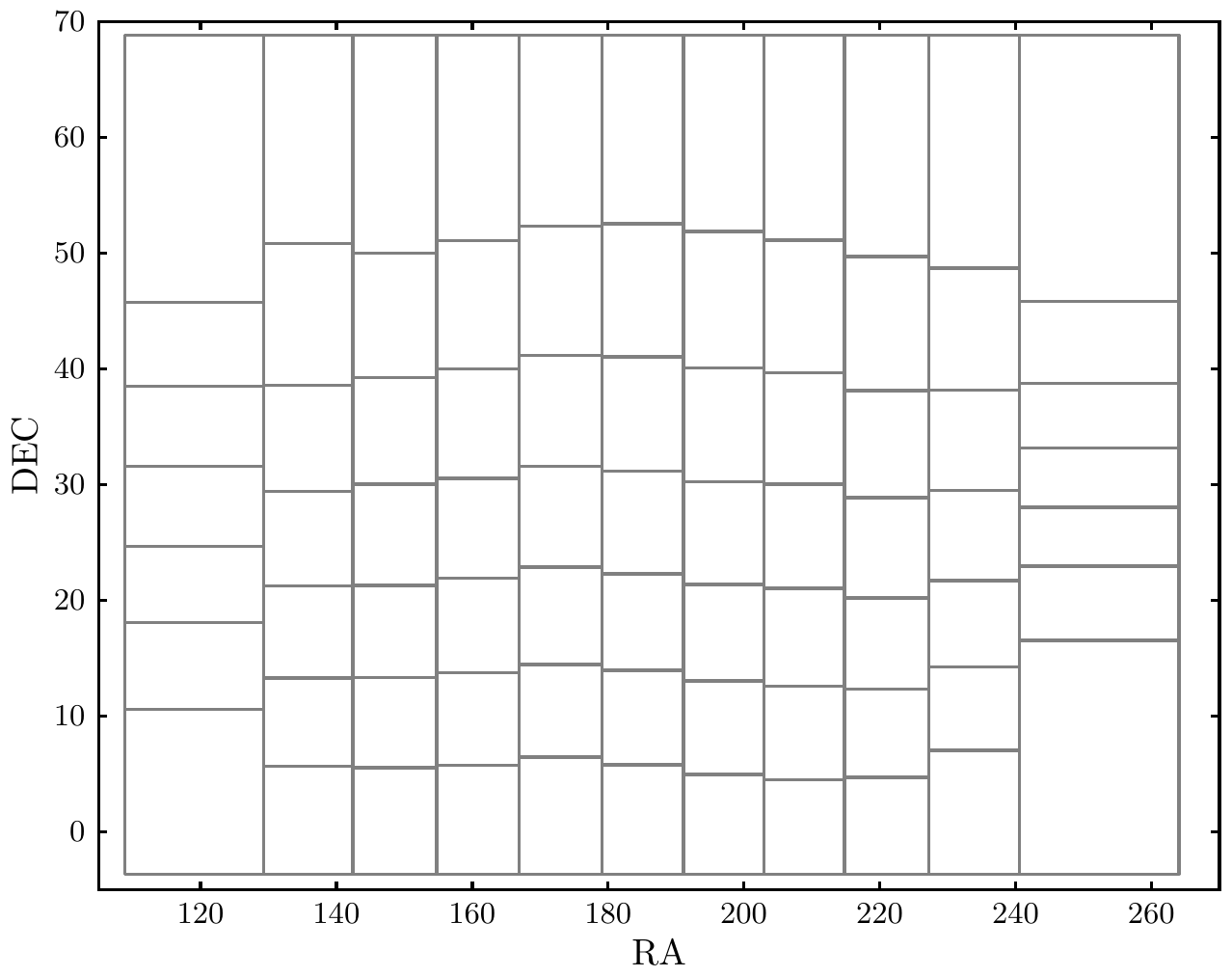}
\end{tabular}
\captionsetup{labelformat=empty}
\caption{\textbf{Supplementary Figure 2:} { \bf Visualization of jackknife regions.} The jackknife regions are chosen so that each region has same effective observed area and is close to square. }
\end{figure}

\begin{figure}
\centering
\begin{tabular}{c}
\includegraphics[scale=0.7]{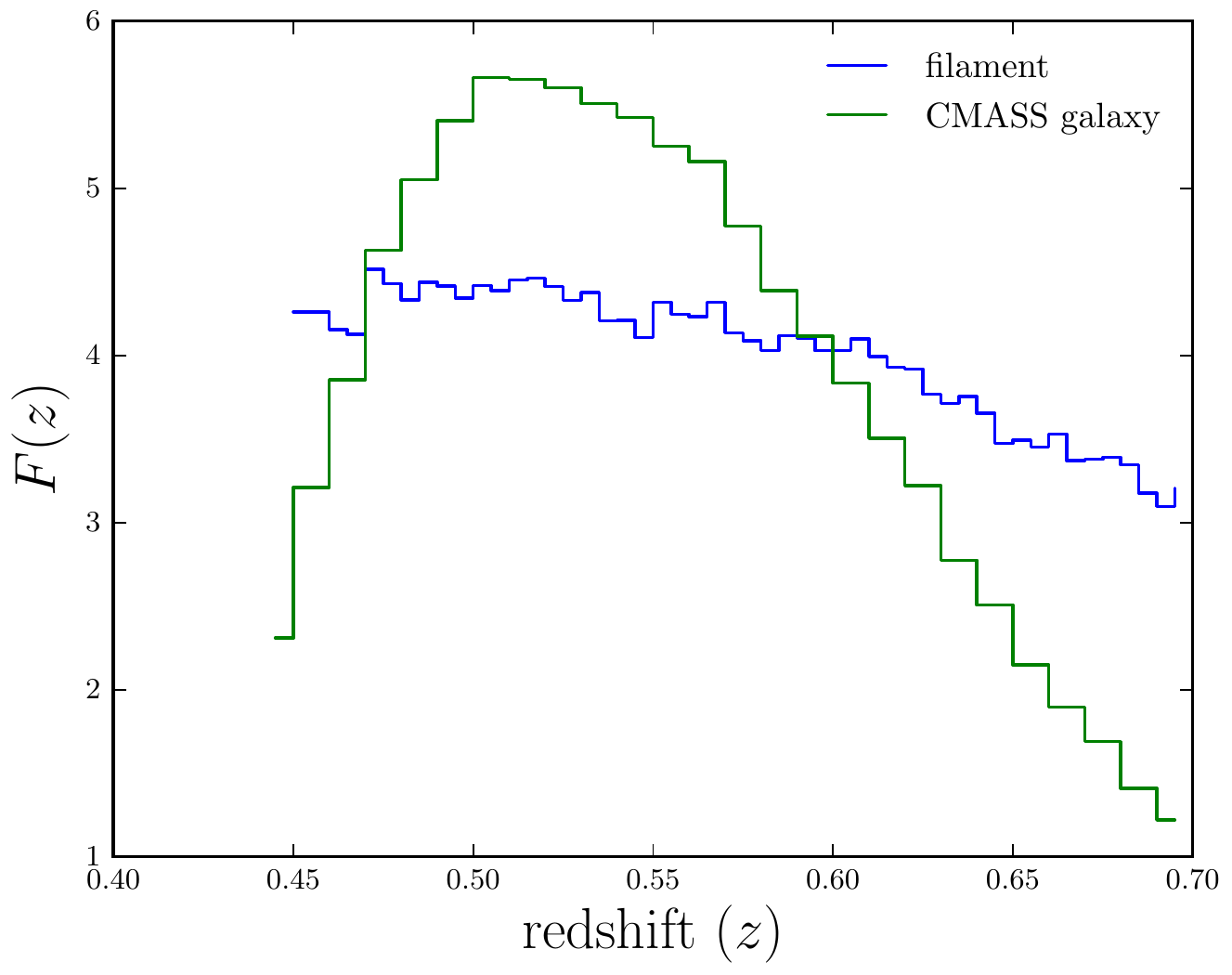}
\end{tabular}
\captionsetup{labelformat=empty}
\caption{\textbf{Supplementary Figure 3:} {\bf Filament intensity distribution and galaxy redshift distribution as a function of redshift.} The blue curve shows the filament intensity distribution as a function of redshift. The green curve shows the CMASS galaxy redshift distribution, defined as the normalized distribution of the number density of galaxies as a function of redshift. The decrease of the filament intensity distribution results from the decrease of CMASS galaxy redshift distribution, from where the filaments in each redshift slice are detected.}
\label{fig:distribution}
\end{figure}

\begin{figure}[h!]
\centering
\begin{tabular}{c}
\includegraphics[scale=0.7]{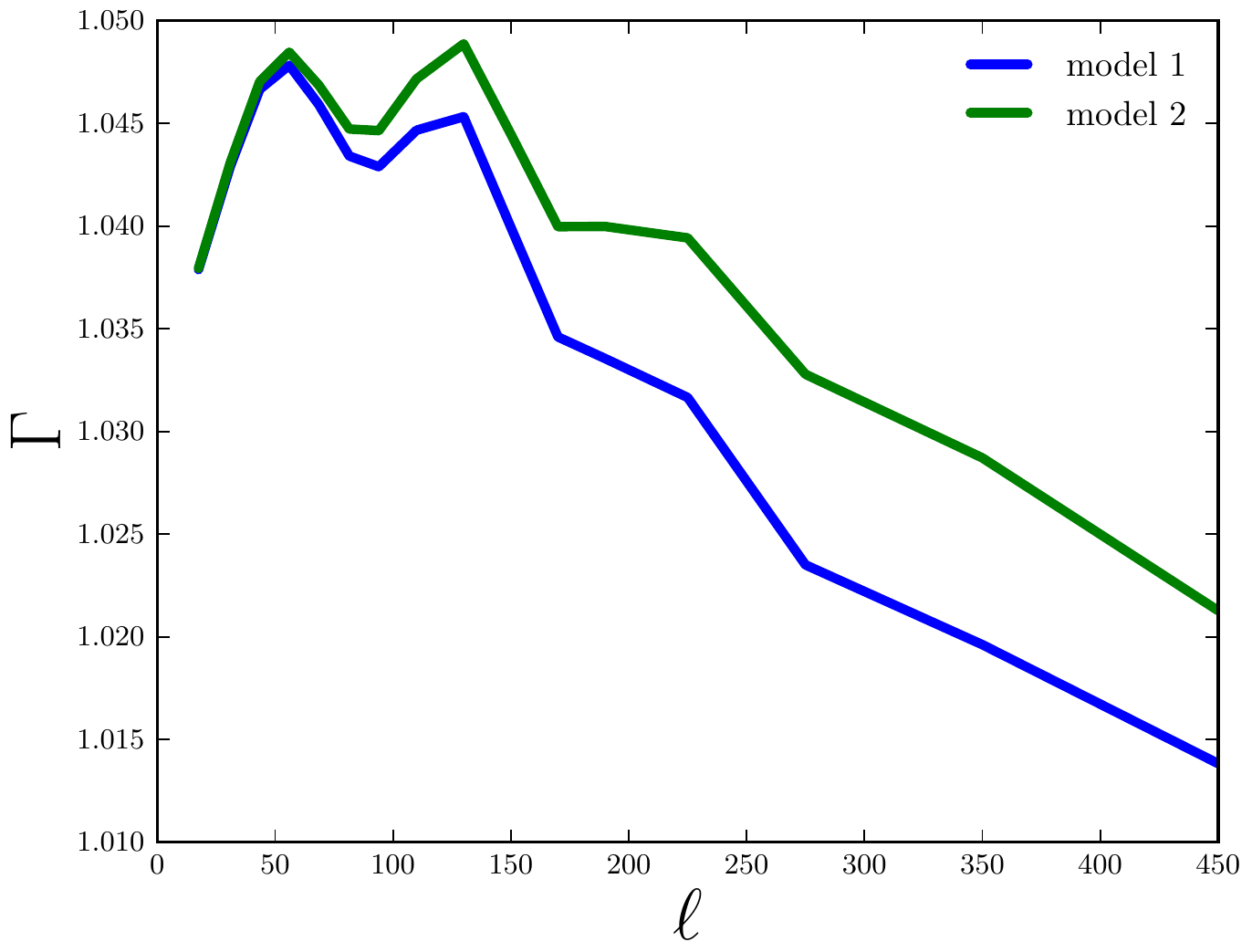}
\end{tabular}
\captionsetup{labelformat=empty}
\caption{\textbf{Supplementary Figure 4:} {\bf Quantification of the deviation between $C_l^{\kappa f}$ estimated from simulation and from theoretical $C_l^{\kappa f}$.} Since the approximations to get $C_l^{\kappa f}$ from simulations are not perfect, the $C_l^{\kappa f}$ from simulation will slightly deviate from the true value of $C_l^{\kappa f}$. $\Gamma$ quantifies the deviation. In theory, We use two models for the smoothing introduced by filaments. In model1, filament length is the overall smoothing scale. In model2, filament length is the smoothing scale along the filament and we fit $\alpha$ for the smoothing in the perpendicular direction, where $1/k_{\perp}(z) \sim \alpha \times 1/k_{\parallel}(z)$ and $k$ is the wavenumber in Fourier space.   As shown in the plot, the deviation of $C_l^{\kappa f}$ between simulation and theory is less than 5\% from unity for both theoretical models.}
\end{figure}

\begin{figure}
\centering
\includegraphics[scale=0.7]{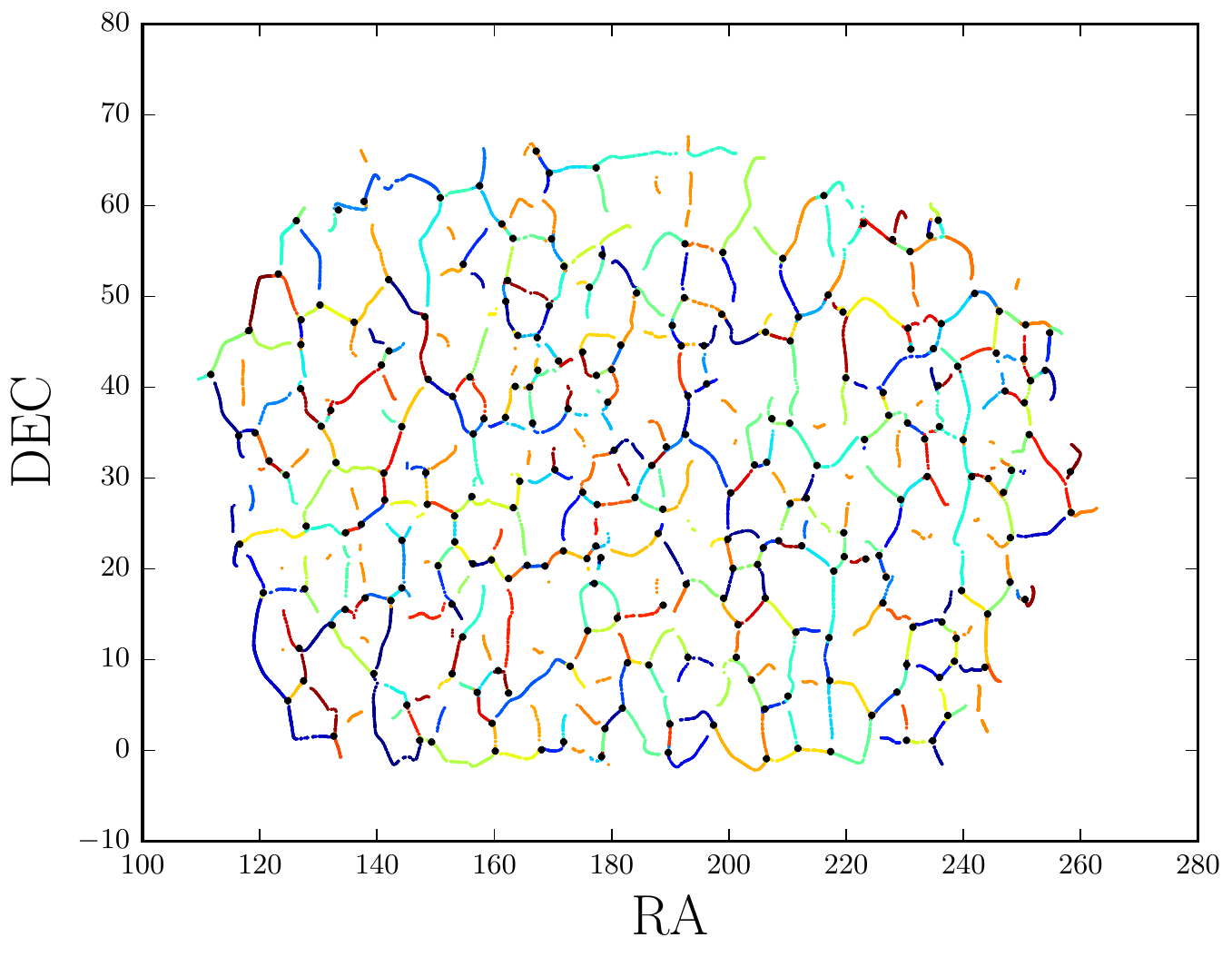}
\captionsetup{labelformat=empty}
\caption{\textbf{Supplementary Figure 5:} {\bf Example of filament grouped in redshift bin 0.55.} A line with the same color is considered as belonging to the same filament.}
\label{fig:fil_length}
\end{figure}

\end{document}